\documentclass[aps,prl,showpacs,twocolumn]{revtex4}

\usepackage[dvips]{graphicx}
\usepackage{epsfig}
\usepackage{amsmath}
\usepackage{amsfonts}
\usepackage{amssymb}
\usepackage{wasysym}

\expandafter\ifx\csname package@font\endcsname\relax\else
 \expandafter\expandafter
 \expandafter\usepackage
 \expandafter\expandafter
  \expandafter{\csname package@font\endcsname}%
\fi

\begin{document}
\def\be{\begin{equation}}
\def\ee{\end{equation}}

\title{On the influence of time and space correlations on the next 
earthquake magnitude}

\author{E. Lippiello$^{a}$, L. de Arcangelis$^{b}$, C. Godano$^{c}$}
%\altaffiliation {
%Department of Physics ``E.R. Caianiello'', University of Salerno,
%84081 Baronissi (SA), Italy}
\affiliation{ $^{a}$ University of Naples ``Federico II'', 80125 Napoli,
Italy\\
$^{b}$ Department of Information Engineering and CNISM, 
Second University of Naples, 81031 Aversa (CE), Italy\\ 
$^{c}$ Department of Environmental Sciences and CNISM, 
Second University of Naples, 81100 Caserta, Italy
}
\begin{abstract}
A crucial point in the debate on feasibility of earthquake prediction is
the dependence of an earthquake magnitude from past seismicity.
Indeed, whilst clustering in time and space is widely accepted, 
much more questionable is the existence of magnitude correlations.
The standard approach generally assumes that magnitudes are independent 
and therefore in principle unpredictable.
Here we show the existence of clustering in magnitude:
earthquakes occur with higher probability close in time, space and 
magnitude to previous events. More precisely, the next earthquake tends to 
have a magnitude similar but smaller than the previous one. 
%This suggests 
%that an earthquake alters the system state in such a way to influence 
%the energy released in subsequent events. 
%The time and space scales of this process appear to be controlled by the 
%magnitude difference between correlated events. 
A dynamical scaling relation 
between magnitude, time and space distances reproduces the 
complex pattern of magnitude, spatial and temporal correlations
observed in experimental seismic catalogs.
\end{abstract}
\maketitle

Since the Omori observation \cite{omori}, temporal clustering is
considered a general and distinct feature of seismic
occurrence. Clustering in space has also been well established
\cite{kagknop} and, together with the Omori law and the Gutenberg-Richter 
law \cite{GR}, is the main ingredient of probabilistic tools for 
time-dependent seismic hazard evaluation \cite{reajon,ogata,gerg}.
The distribution $D(\Delta t)$  of the inter-time $\Delta t$  elapsed
between 
two successive events is a 
suitable quantity  to characterize the temporal
organization of seismicity. Analogously, the distribution $D(\Delta
r)$ of the distance $\Delta r$ between subsequent epicenters provides useful
insights in the spatial organization. Both distributions have been
the subject of much interest in the last years
\cite{bak,mega,yang,corral,corral1,scaf,lind,livina,noisoc,sorn2,davpac,corral2,corral3,noi}. 
In particular, they exhibit universal behavior 
essentially independent of the space region and the magnitude range
considered \cite{corral,sorn2,davpac,corral3}.
Furthermore, the question  of the existence of correlations between 
magnitudes of subsequent earthquakes 
has been also recently addressed \cite{noi,corral1,corral2}. 
In ref.\cite{corral1,corral2}, Corral has shown that the
Southern  California Catalog exhibits possible
magnitude correlations that are small but different from
zero. However, restricting his investigation to earthquakes with
$\Delta t$ greater then $30$ minutes, he observes that correlations reduce
and become smaller than statistical uncertainty. 
Magnitude correlations have
been, therefore, interpreted as a spurious effect due to short term
aftershock incompleteness (STAI) \cite{kagan}. According to this hypothesis, 
aftershocks, in particular small events, occurring closely after large shocks
are not reported in the experimental catalog. This interpretation
agrees with the standard approach
that assumes independence of earthquake magnitudes:
an earthquake "does not know how large it will become". This has
strong implications on the still open question of  
earthquake predictability \cite{nat}. 
On the other hand, a recent analysis of the Southern California
Catalog \cite{noi} 
has shown the existence of non-zero magnitude correlations, not to be 
attributed to STAI. These are observed by means of an
averaging procedure that reduces statistical
fluctuations.  A dynamical scaling hypothesis
relating magnitude to time differences has been proposed to explain 
the observed magnitude correlations.

In this paper we 
present a statistical analysis of experimental catalogs that
confirms the existence of relevant magnitude correlations.
In particular, the analysis enlightens the structure of these
correlations 
and their relationship with $\Delta t$ and $\Delta r$. 
We then introduce a trigger model based on a  
dynamical scaling relation between energy, space and time and 
show that this model reproduces the above experimental findings.  
We consider the NCEDC catalog (downloaded at 
http:///www.ncedc.org/ncedc/, years 1974-2002,
South (North) lat. 32 (37), West (East) long. -122 (-114)). 
Similar results are obtained for seismic catalogs of other
geographic regions.
To ensure catalog completeness, we consider only events with 
magnitudes $m \ge 3$ and take into account
STAI using the method proposed in ref.\cite{helmkag}.
The quantities considered are 
$\Delta t_i=t_{i+1}-t_i$,
$\Delta r_i =\vert \vec r_{i+1}-\vec r_i\vert $ and $\Delta m_i=m_{i+1}-m_i$,
i.e. the time, space and magnitude difference between subsequent
 events. We have also evaluated the quantity $\Delta
 m^*_i=m_{i^*}-m_i$ where $i^* \neq i$ is the random index of an
 earthquake recorded in the catalog. Hence, $\Delta m_i^*$ is the magnitude
 difference within a reshuffled catalog where the magnitude of the
 subsequent earthquake is independent of previous ones. We then
 consider the conditional probability 
\be
P(\Delta x_i<x_0 \vert \Delta
 y_i<y_0) \equiv \frac{N(x_0,y_0)}{N(y_0)}
\label{pcond}
\ee
 where $N(x_0,y_0)$ is the number of couples of subsequent events
 with both $\Delta x_i <x_0$ and $\Delta y_i <y_0$ and $N(y_0)$ is the
 number of couples with $\Delta y_i <y_0$. In the following
 $\Delta x_i$ or $\Delta y_i$ will be used to indicate, depending on
cases,  $\Delta r_i$, $\Delta
 t_i$, $\Delta m_i$ or $\Delta m_i^*$.       
Our method is schematically presented in Fig.1.
Keeping  $m_0$ and $r_0$ fixed, we compute the quantity 
$P(\Delta  m^*_i<m_0 \vert \Delta r_i<r_0)$ for several independent
random realizations of the reshuffled catalog, obtaining the distribution
$\rho \left [P(\Delta
 m^*_i<m_0 \vert \Delta r_i<r_0)\right ]$. 
Taking $10^4$ independent realizations of the magnitude reshuffling,
for each given $m_0$ and $r_0$, we always find that
 $\rho \left [P(\Delta m^*_{i}<m_0 \vert \Delta r_
 {i} <r_0)\right ]$ is
 gaussian distributed with mean value $Q(m_0,r_0)$ 
 and standard deviation $\sigma(m_0,r_0)$. Analogous behaviour is obtained
for $P(\Delta  m^*_i<m_0 \vert \Delta t_i<t_0)$ and we similarly define
$Q(m_0,t_0)$ and $\sigma(m_0,t_0)$. 
The relevant quantity is  $\delta P(m_0,y_0)=P(\Delta m_{i}<m_0
 \vert \Delta y_{i} <y_0)-Q(m_0,y_0)$, i.e the difference between the
 value of $P(\Delta m_{i}<m_0 \vert \Delta y_
 {i} <y_0)$ in the real catalog and its mean value in the reshuffled one.
If  the absolute value $\vert \delta P(m_0,y_0) \vert$ is larger than  
$\sigma (m_0,y_0)$,  significant non-zero correlations between
magnitudes of successive earthquakes exist. 
In particular, a positive value of
$\delta P(m_0,y_0) > \sigma(m_0,y_0)$ implies that the  
number of couples $N(m_0,y_0)$ is significantly larger in the
 real catalog with respect to a catalog where magnitudes are
 uncorrelated. 
In Fig.1 we explicitly compare $\rho \left [P(\Delta
 m^*_i<m_0 \vert \Delta y_i<y_0)\right ]$ with $P(\Delta
 m_i<m_0 \vert \Delta y_i<y_0)$ for $m_0=0$ and $y_0=r_0=10 km$
or $y_0=t_0=1h$. One clearly observes the existence of non-zero
 magnitude correlations, since
$\delta P(m_0,r_0) \simeq 8.3 \sigma(m_0,r_0)$  and     
$\delta P(m_0,t_0) \simeq 7.3 \sigma(m_0,t_0)$.    
For a deeper understanding of the nature of the observed correlations,
 the above analysis has been extended to other values of $m_0$, $r_0$
 and $t_0$. 
 In Fig.2 and Fig.3 we plot the quantities
 $\delta P(m_0,r_0) $ and $\delta P(m_0,t_0)$  as a function of $m_0$  for
 different values of $r_0$ and $t_0$ respectively. The error bar of
 each point is the standard deviation $\sigma(m_0,y_0)$.
We first observe that for each value of $r_0$ and $t_0$ and for a wide
 range of $m_0$, $\delta P(m_0,y_0)$ is strictly positive and
 significantly different from zero. Considering the behavior at fixed
 $r_0$ or $t_0$, the 
curve has a peak centered in $m_0 \lesssim
 0$, indicating a crossover from positive to negative
 correlations. 
This can be better enlightened by the derivative
 $P^\prime (m_0,y_0)=\frac{d \delta P(m_0,y_0)}{d m_0}=$,
which represents the probability difference for $\Delta m_i=m_0$
 conditioned to $\Delta y_i<y_0$ . $P^\prime(m_0,y_0)$ is therefore an
 estimate of the magnitude correlation between two  subsequent events 
with $\Delta m_i=m_0$. 
 Interestingly, for both $y_0=r_0$ and $y_0=t_0$ (inset of Fig.2 and Fig.3), 
$P ^\prime (m_0,y_0)$ has the maximum value for $m_0$ in
 the range $[-1,-0.5]$ and decreases to zero for smaller values of $m_0$. 
For $m_0 \ge 0$, $P^\prime (m_0,y_0)$ is always  negative with the minimum
value centered around $m_0 \in [0,0.5]$ and going to zero for large
$m_0$. This implies that, for positive $m_0$, 
the probability is larger in a reshuffled catalog
than in the real one.
As a consequence, Fig.s (2,3) clearly show that the
 magnitudes of subsequent earthquakes are correlated and, 
in particular, the next earthquake tends to 
have a magnitude close but smaller than the previous one.
Furthermore, Fig.s (2,3) indicate that 
for any fixed $m_0$, curves corresponding to different $r_0$ or $t_0$
 clearly separate, showing the existence of correlations
 between $\Delta m$ and $\Delta r$, $\Delta m$ and $\Delta t$.
In particular we observe that the 
larger are $r_0$ or $t_0$, the
 smaller are magnitude correlations.

To better investigate the role of $r_0$ and $t_0$ on magnitude
correlations, we consider $P(\Delta r_i<r_0 \vert \Delta
 m_i<m_0)$ and $P(\Delta t_i<t_0 \vert \Delta m_i<m_0)$.
Following the procedure described for Fig.1, we  compute 
$\delta P(x_0,m_0)$ and $\sigma(x_0,m_0)$ for $x_0=r_0,t_0$ (Fig.4). 
Also in this case, a non-zero $\delta P(x_0,m_0)$ is the
signature of magnitude correlations. We observe that, 
for each value of $m_0$, $\delta P(x_0,m_0)$ is a decreasing
function of $r_0$ and $t_0$ and therefore stronger correlations
are observed for events that occur closely in time and space. More
specifically, Fig.4 shows that for smaller values of
$t_0$ and $r_0$, the probability to have $\Delta m_i \le -2$ is about
$40\%$ larger in the real than in a given reshuffled catalog.

\begin{figure}
\includegraphics[width=8cm]{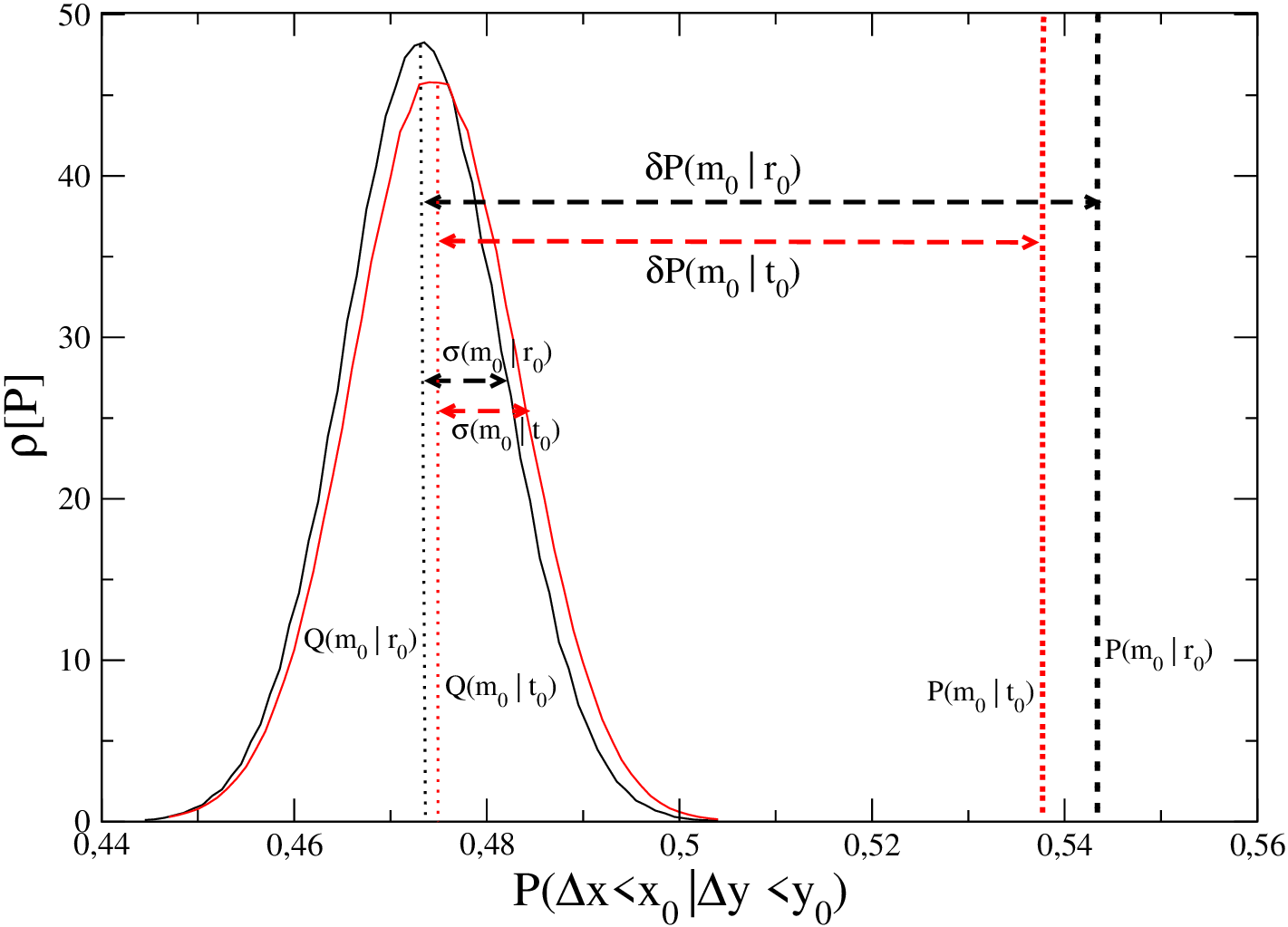}% Here is how to import EPS art
\caption{(Color online) 
The distribution of  $P(\Delta m^*_{i}<m_0 \vert \Delta r_
 {i} <r_0)$  for $r_0=10 Km$ and
 $m_0=0$ (black curve) is compared with  $P(\Delta m_{i}<0 \vert \Delta r_
 {i} <10 Km)=0.543$  (broken black curve). $\rho[P]$ has a gaussian
 behaviour with mean $Q(m_0,r_0)=0.473$ and standard deviation 
 $\sigma(m_0,r_0)=0.00825$. For $r_0=10 Km$ and $m_0=0$ one has $\delta
 P(m_0,r_0)=0.07 \simeq 8.5 \sigma(m_0,r_0)$ strongly supporting the
 existence of correlations between $m_i$ and $m_{i-1}$. The same
 conclusion can be obtained by considering $P(\Delta m^*_{i}<m_0 \vert 
\Delta t_{i} <t_0)$ for $t_0=1 h$ and
 $m_0=0$ (red curve). It is found $P(\Delta m_{i}<0 \vert \Delta t_
 {i} <1 h)=0.537$  (broken red curve) whereas  $\rho[P]$ is a
 gaussian with mean $Q(m_0,t_0)=0.475$ and standard deviation
$\sigma(m_0,t_0)=0.0085$. As a consequence
 $\delta P(m_0,t_0) \simeq 7.3 \sigma (m_0,t_0)$.    
\\}
\end{figure}

\begin{figure}
\includegraphics[width=8cm]{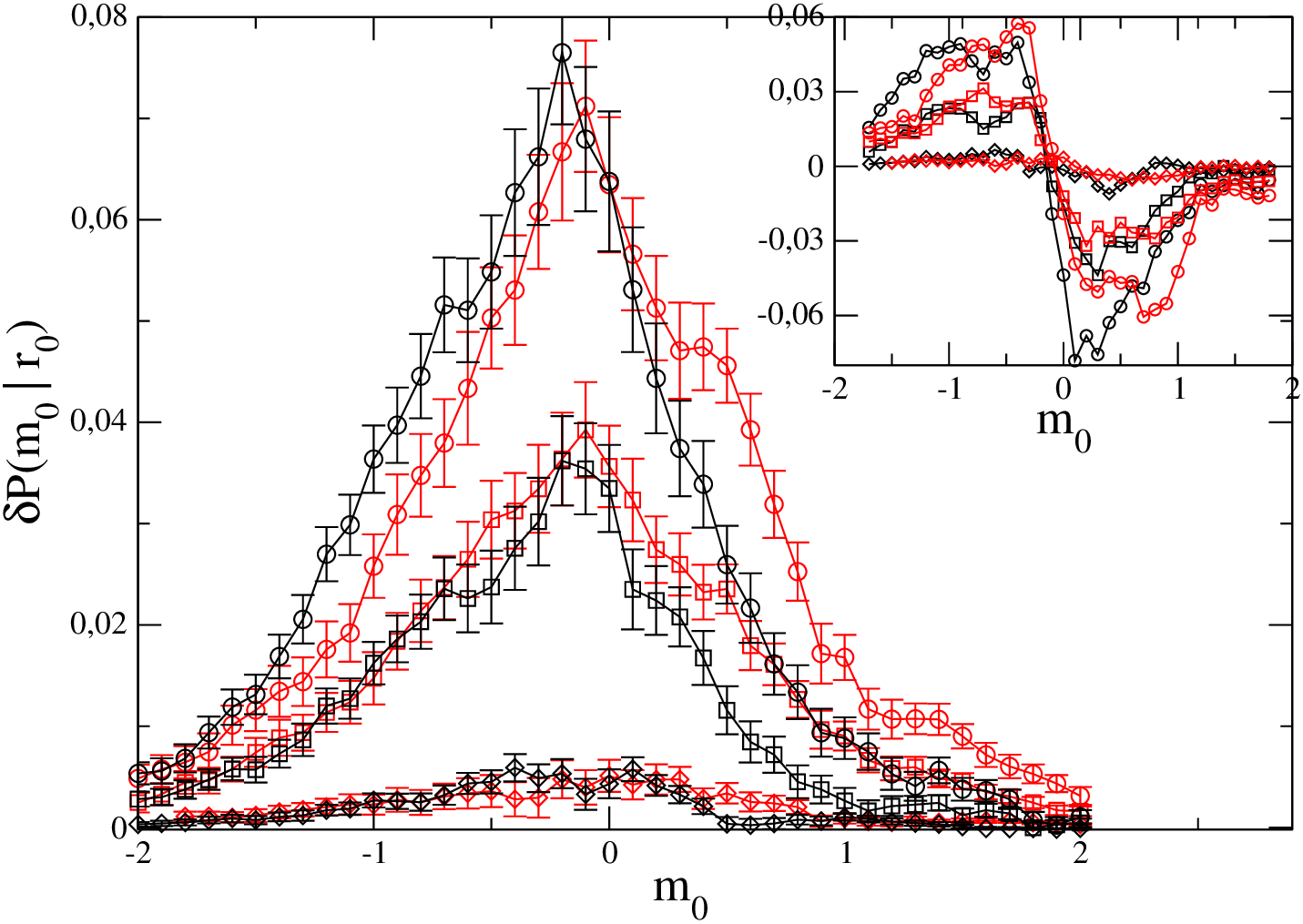}% Here is how to import EPS art
\caption{(Color online) 
The quantity $\delta P(m_0,r_0)$ as a function
  of $m_0$ for $r_0=10, 100, 500 Km$ from top to bottom. For each $r_0$ and
  $m_0$ the error bar is the standard deviation
  $\sigma(m_0,r_0)$. Data for the Southern California catalog
(black) are compared with numerical simulations (red). In the inset, the
  quantity $P^\prime(m_0,r_0)$.\\      
\\}
\end{figure}

\begin{figure}
\includegraphics[width=8cm]{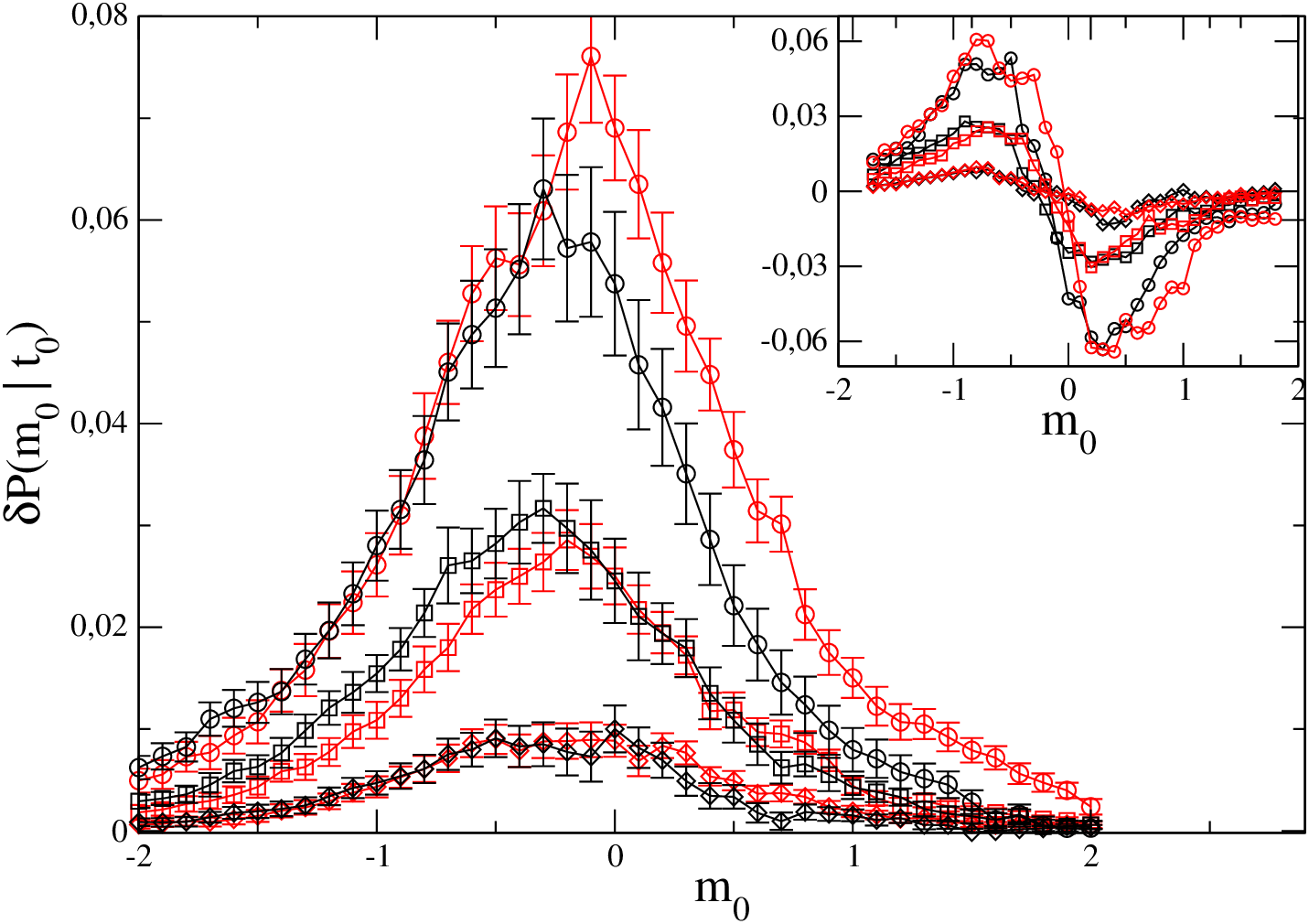}% Here is how to import EPS art
\caption{(Color online)
The quantity $\delta P(m_0,t_0)$ as a function
  of $m_0$ for $t_0=1, 10, 50 hours$  from top to bottom. For each $t_0$ and
  $m_0$ the error bar is the standard deviation $\sigma(m_0,t_0)$. 
Data for the Southern California catalog
(black) are compared with numerical simulations (red). In the inset, the
  quantity $P^\prime(m_0,t_0)$.
\\ }      
\end{figure}

\begin{figure}
\includegraphics[width=8cm]{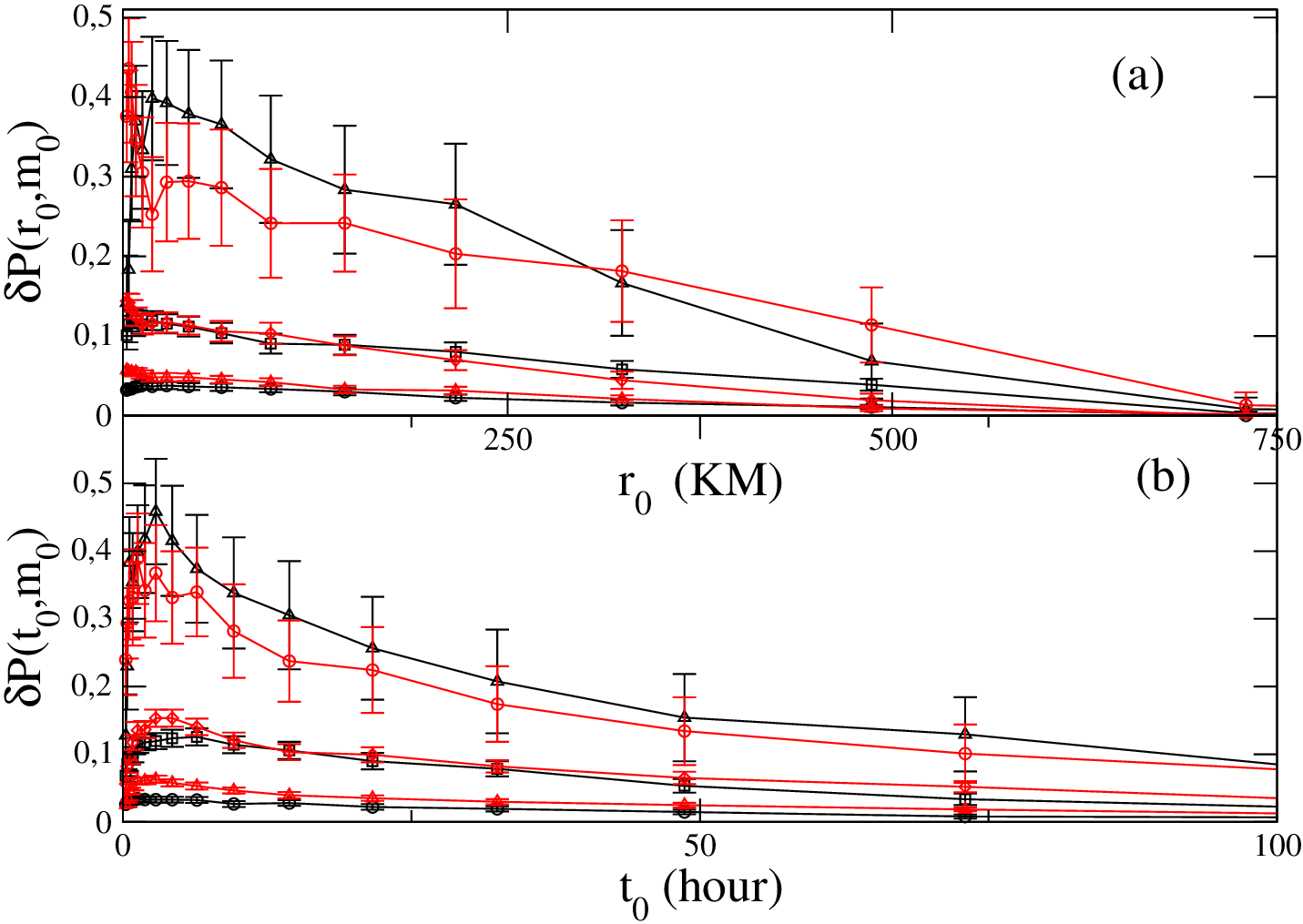}% Here is how to import EPS art
\caption{(Color online) 
(a)  The quantity $\delta P(r_0,m_0)$ as a function
  of $r_0$ for $m_0=2, 0.5, 0$ from top to bottom.
(b)   The quantity $\delta P(t_0,m_0)$ as a function
  of $r_0$ for $m_0=2, 0.5, 0$ from top to bottom. 
Data for the experimental catalog
(black) are compared with numerical simulations (red).
For each point the error bar is the standard deviation $\sigma(x_0,m_0)$.  
\\ }
\end{figure}

%\begin{figure}
%\includegraphics[width=8cm]{fig5a.eps}% Here is how to import EPS art
%\caption{(Color online) 
%  The quantity $\delta P(t_0,m_0)$ as a function
%  of $r_0$ for $m_0=2, 0.5, 0$ from top to bottom. 
%\\}
%\end{figure}

The above analysis shows that a better description of real seismicity
can be obtained if correlations between time, space and magnitude are
properly taken into account. As in dynamical critical phenomena where
energy and time fix a characteristic length scale, similar ideas can
be used to introduce magnitude correlations within standard trigger
models for seismicity. In trigger models \cite{vere-jones}, the 
probability  to have the next earthquake in
the time window $[t,t+\delta t]$, with epicenter in the region $[\vec
  r, \vec r+\delta \vec r]$ and magnitude in the range $[m,m+\delta
  m]$ is given by the superposition 
\be
{\cal P}(t,\vec r,m)=\sum_{j}P(t-t_j,\vert \vec r -\vec r_j \vert ,m,m_j) ,
\label{trigger}
\ee 
where $P(t-t_j,\vert \vec r -\vec r_j\vert  ,m,m_j)$ is the 
probability conditioned to the occurrence of an earthquake of
magnitude $m_j$, at time $t_j<t$, in the position $\vec r_j$ .
In the widely accepted ETAS model \cite{ogata},
$m,m_j,t-t_j$ and $\vert \vec r -\vec r_j \vert$ are all independent
quantities and empirical laws are used to characterize their
distributions.  Many analytical and numerical studies show that the
ETAS model captures several aspects of real seismic
occurrence \cite{ogata,ogata2,helmstetter,etas,sorn2,saichev1}. 
Nevertheless, because of the assumption of independence between $m$ and $m_j$, 
$\delta P(x,y)$ would be a random fluctuating
function with zero average and standard deviation
$\sigma(x,y)$, in all cases considered in Fig.s (2-4). 
Hence, by construction, the ETAS model does not take
into account magnitude correlations and their dependence on time and space.

In order to reproduce the experimental findings,  we introduce
\be
\tau_{ij}=k_t 10^{b(m_i-m_j)}  \hskip0.5cm {\rm and}\hskip0.5cm
r_{ij}^z=k_r \vert \vec r_i-\vec r_j \vert ^z
\label{taurij} 
\ee
which fix two characteristic 
time scales leading to the scaling behaviour with $\Delta t_{ij}=t_i-t_j$
\be
P(\Delta t_{ij},\vert \vec r_i -\vec r_j \vert,m_i,m_j)=\Delta t_{ij}^{-2/z}
H\left ( \frac{\tau_{ij}}{\Delta t_{ij}},
\frac {r_{ij}}{\Delta t_{ij}^{1/z}} \right ). 
\label{scalingtot}
\ee
The exponent $2/z$ is determined by
imposing the condition $\int d \vec r_i P(\Delta t_{ij},\vert \vec r_i -\vec r_j
\vert,m_i,m_j)=H_1\left ( \frac{\tau_{ij}}{\Delta t_{ij}}\right )$, where
the function $H_1(x)$ must satisfy the normalization condition 
$ \int dx H_1(x) =1$. Following ref. \cite{noi} it is possible
to show that this normalization removes the problem with
``ultaviolet'' and ``infrared'' divergences of the ETAS model
\cite{saichev1}.  

In order to simplify the numerical procedure, we 
consider a special case of Eq.(\ref{scalingtot})  
\be
H\left ( \frac{\tau_{ij}}{\Delta t_{ij}},
\frac{r_{ij}}{\Delta t_{ij}^{1/z}} \right )  
=
H_1\left ( \frac{\tau_{ij}}{\Delta t_{ij}} \right) H_2 
\left(\frac{r_{ij}}{\Delta t_{ij}^{1/z}} \right ). 
\label{scalingtot2}
\ee
In the numerical simulation, we generate a synthetic catalog
containing only occurrence times and magnitudes. We start with a random event 
at initial time $t_0=0$, time is then increased by one unit and 
a trial magnitude is randomly chosen.
The $i$-th event, at time $t_i$ and with magnitude $m_i$,
occurs with a probability  
$\sum _{j<i} H_1(\tau _{ij}/\Delta t_{ij})$ where the sum is over all previous
events. In particular we use
\be
   H_1(x)=\frac{A}{e^{x}-1+\gamma_1}
\label{h1}
\ee
with the parameters
$k_t=12.7 h$, $A=0.21 h^{-1}$, $\gamma_1=0.1$ and $b=0.9$,
 that in ref.\cite{noi} are found able to
reproduce the experimental behavior of magnitude and inter-time
distributions. To introduce the epicenter location in the numerical
catalog we use the power law 
\be
H_2(x) =\frac{B}{  x^{\mu}+\gamma_2}
\label{h2}
\ee
where $\mu$ and $\gamma_2$ are fit-parameters and $B$ is fixed by the
normalization.
Other functional forms for $H_2(x)$ give similar results but with a worse
agreement with experimental data. 
We follow the method used in ref.\cite{helmstetter,ogata2} for the ETAS model.
More precisely, for  $j=0$ the epicenter of the first event in the
catalog $\vec r_{j=0}$ 
is randomly fixed in a point of a square lattice of size $L$.
For sufficiently large lattices, the results are $L$ independent.
Next, $j$ is updated $j =j+1$, and the
mother of the $j-th$ earthquake is chosen among all previous $0 \le i
\le j-1$
events according to the probability $H_1\left (
\frac{\tau_{ij}}{\Delta t_{ij}} \right)$. Once the mother event is
identified, its epicenter  $\vec r^*$ and occurrence time $t^*$ are 
used to randomly obtain $\vec r_j$ from the probability
distribution 
$(t_j-t^*)^{-2/z}H_2 \left(\frac{\vert \vec r_{j}-\vec r^*
  \vert}{(t_j-t^*)^{1/z}}\right )$ and
assuming space isotropy.

The numerical catalog is analyzed 
with the same procedure applied to experimental data.
Numerical results for $\delta P(x_0,y_0)$
are presented as red circles in Fig.s (2-4) for $\mu=2.6$, $k_r=0.03 h/Km^z$, $z=3.3$ 
and $\gamma_2=0.1$. We observe a very
good agreement between numerical and experimental data. In particular,
$\delta P(m_0,y_0)$ (Fig.s(2,3)) displays a maximum value
localized around the maximum of the experimental
distribution. Furthermore, also the functional form of the decay of
$\delta P(x_0,m_0)$ for both $x_0 =r_0$ (Fig.4a) and $x_0=t_0$ (Fig.4b) is
reproduced.

The agreement between numerical and experimental results, indicates
that the scaling relation (\ref{scalingtot}) among magnitudes, times,
and epicenter distances can describe the complex pattern of
the experimentally observed correlations. The origin of magnitude
correlations, within our theoretical
approach, has a direct interpretation. According to Eq.s
(\ref{scalingtot2},\ref{h1},\ref{h2}), indeed, at the time $t$ an
earthquake of magnitude $m$ with epicenter $\vec r$ has a finite
probability to be triggered by a previous ($m_j,t_j,\vec r_j$)
earthquake only if $m < m_j -(1/b)\log((t-t_j)/k_t)$ and 
$r_{ij} < (t-t_j)^{1/z}$ (Eq. (\ref{taurij})). As a
consequence, only events occurring close in time and space can have a
magnitude close and smaller, or even larger, that the previous triggering
one. Magnitude correlations, therefore, become particularly relevant
within aftershock sequences, when earthquakes tend to be very close in
time and space. Thus a dynamical scaling approach, that properly takes into
account these correlations, can improve existing methods for time dependent
hazard evaluation.


\begin{thebibliography}{}

\bibitem{omori} F. Omori, {\it J. Coll. Sci. Imp. Univ.
Tokyo} {\bf 7}, 111, (1894)

\bibitem{kagknop} 
Y.Y. Kagan, L. Knopoff, {\it Geophys. J. Roy. Astron. Soc.} {\bf 62},
303 (1980)

\bibitem{GR} B. Gutenberg,
C.F. Richter, {\it Bull. Seism. Soc. Am.} {\bf 34}, 185 (1944)

\bibitem{ogata} Y. Ogata, {\it J. Amer. Stat. Assoc.}
{\bf 83}, 9, (1988)

\bibitem{reajon}P.A Reasenberg,  L.M. Jones, {\it Science}
{\bf 243}, 1173 (1989)

\bibitem{gerg} M.C. Gerstenberger, S. Wiemer,
L.M. Jones, P.A. Reasenberg, {\it Nature} {\bf 435}, 328 (2005)


\bibitem{bak} P. Bak, K. Christensen, L. Danon, T. Scanlon, 
{\it Phys. Rev. Lett.} {\bf 88}, 178501, (2002)

\bibitem{mega} M.S. Mega {\it et al.},
{\it Phys. Rev. Lett.} {\bf 90}, 188501 (2003)

\bibitem{corral} A. Corral, {\it Phys. Rev. Lett.} {\bf 92}, 108501
  (2004)

\bibitem{yang}  X. Yang, S. Du,  J. Ma, {\it Phys. Rev. Lett.}
{\bf 92}, 228501 (2004)

\bibitem{scaf} N. Scafetta, B.J.
West {\it Phys. Rev. Lett.} {\bf 92}, 138501 (2004).

\bibitem{lind} M. Lindman, K. Jonsdottir, R. Roberts, B. Lund, R. Bodvarsson,
 {\it Phys. Rev. Lett.} {\bf 94},
 108501 (2005)

\bibitem{livina} 
V. N. Livina, S. Havlin,  A. Bunde
{\it Phys. Rev. Lett.} {\bf 95}, 208501 (2005)

\bibitem{noisoc} E. Lippiello, C. Godano, L. de Arcangelis, {\it
  Europhys. Lett.} 
{\bf 72}, 678 (2005)

\bibitem{sorn2} A. Saichev, D. Sornette, {\it Phys. Rev. Lett.} {\bf 97},
078501 (2006). 

\bibitem{davpac} J. Davidsen, M. Paczuski, {\it Phys. Rev. Lett.} {\bf 94},
  048501 (2005). 

\bibitem{corral3}  A. Corral, {\it Phys. Rev. Lett.} {\bf 97}, 178501 (2006) 

\bibitem{corral1} A. Corral, {\it Phys. Rev. Lett.} {\bf 95}, 159801 (2005) 

\bibitem{corral2}A. Corral, {\it TectonoPhysics} {\bf 424}, 177 (2006)


\bibitem{noi}E. Lippiello, C. Godano, L. de Arcangelis, 
{\it Phys. Rev. Lett.} 
{\bf 98}, 098501 (2007)

\bibitem{kagan} Y.Y. Kagan, {\it Bull. Seism. Soc. Amer.}, {\bf 94(4)}, 1207 
(2004)

\bibitem{nat} 
http://www.nature.com/nature/debates/earthquake/ index.html 

\bibitem{helmkag} A. Helmstetter, Y. Kagan, D. Jackson, 
{\it J. Geophys. Res.} {\bf 110}, B05S08 (2005)

\bibitem{vere-jones} J. F. D. Vere-Jones, {\it J. Roy. Statist. Soc.},
 {\bf B32}, 1, (1970)


\bibitem{ogata2} Y. Ogata, {\it Ann. Inst. Stat. Math}
{\bf 50}, 379, (1998)

\bibitem{helmstetter}
A. Helmstetter, D. Sornette,
{\it Phys. Rev. E}  {\bf66} 061104 1, (2002);

\bibitem{etas} 
A. Helmstetter, D. Sornette, {\it J. Geophys. Res.}  {\bf107} 2237, (2002);
A. Saichev, D. Sornette {\it Phys. Rev. E}, {\bf 70}, 046123
(2004); A. Saichev, A. Helmstetter, D. Sornette, {\it Pure and
Applied Geophysics}, {\bf 162}, 1113, (2005);
D. Sornette, M.J. Werner
{\it J. Geophys. Res.}, {\bf 110}, B09303, (2005)  

\bibitem{saichev1} A. Saichev, D. Sornette, {\it Phys. Rev. E} {\bf 72}, 
056122 (2005)




\end{thebibliography}
\end{document}